\documentstyle[pra,aps,epsf,floats]{revtex}
\begin{document}
\twocolumn
\draft
\title{Quantum channel locally interacting with environment
}
\author{Satoshi Ishizaka\cite{Ishizaka}}
\address{Fundamental Research Laboratories,\\ 
System Devices and Fundamental Research, 
NEC Corporation,\\
34 Miyukigaoka, Tsukuba, Ibaraki, 305-8501, Japan}
\date{\today}
\maketitle
\begin{abstract}
The quantum channel subject to local interaction with two-level environment
is studied.
The two-level environment is regarded as a quantum bit (qubit) as well as
a pair of particles owned by Alice and Bob.
The amount of entanglement initially shared by Alice
and Bob is distributed among these three qubits due to the interaction. 
In this model, we show that the singlet fraction of the decohered
quantum channel is uniquely determined by the distributed entanglement.
When the decohered quantum channel is used under the standard teleportation
scheme, the optimal teleportation fidelity is well understood by considering
the remaining entanglement between environment and transmitted state.
\end{abstract}
\narrowtext
Entanglement is an important resource for most applications of quantum
information, and a number of measures quantifying the amount of entanglement,
such as the entanglement of formation \cite{Bennett96a}, have been
proposed.
In the quantum teleportation \cite{Bennett95b},
two bits of classical information and a pair of particles in a maximally
entangled state can transmit an unknown quantum state faithfully.
The entangled particles owned by Alice and Bob act as a quantum
channel carrying quantum information.
In the {\it standard teleportation scheme} (STS) \cite{Bennett95b,Badziag00a},
Horodecki showed that the optimal fidelity of the teleported state
is given by \cite{Horodecki99a}
\begin{equation}
f=\frac{2F_{AB}+1}{3},
\label{eq: Fidelity}
\end{equation}
where $F_{AB}$ is the singlet fraction 
of the channel state $\varrho_{AB}$ in the bipartite $2\times2$ system.
The singlet fraction is defined
as $F_{AB}\!=\!\max \langle e|\varrho_{AB}|e \rangle$,
where the maximum is taken over the all maximally entangled states $|e\rangle$
\cite{Bennett96a}.
\par
When the quantum channel is a pure state, the singlet fraction can be
regarded as a measure of entanglement.
In fact, the singlet fraction is related to the
Hill-Wootters concurrence \cite{Hill97a,Wootters98a} as
$F_{AB}\!=\!(1+C_{AB})/2$.
Therefore, the optimal fidelity in the STS is written by the concurrence
as
\begin{equation}
f=\frac{2}{3}+\frac{1}{3}C_{AB}.
\label{eq: Pure State Fidelity}
\end{equation}
However, when the quantum channel is a mixed state, the singlet fraction is
no longer the measure of entanglement.
The entanglement does not increase under the local quantum operations and
classical communications, but the singlet fraction does.
In fact, recently Badziag, Horodecki \cite{Badziag00a}, and Bandyopadhyay
\cite{Bandyopadhyay00a} showed this case.
\par
When the quantum channel interacts with the surrounding environment,
the quantum channel is entangled with the environment and falls into the
mixed state.
When such the decohered quantum channel is used for the quantum teleportation,
the teleported state is generally entangled with the environment, and
some relation between the remaining entanglement and teleportation fidelity
will be expected.
\par
In this report, we shall investigate the quantum channel subject to local
interaction with environment.
We restrict ourselves to the case that the environment is a two-level system
for simplicity.
Therefore, the environment is regarded as a quantum bit (qubit) as well
as a pair of particles owned by Alice and Bob, and three qubits constitute
the total system.
In this model, the amount of entanglement initially shared by Alice and Bob
is distributed among these three qubits due to the interaction.
We show that the singlet fraction of the decohered quantum channel
is uniquely determined by the distributed entanglement.
When the decohered quantum channel is used under the STS, the optimal
teleportation fidelity is well understood by considering
the remaining entanglement between transmitted state and environment.
\par
Let Alice and Bob initially share an ideal (perfect) quantum channel: a pair of
particles in the maximally entangled state of 
$|\phi^+\rangle\!=\! (|00\rangle+|11\rangle)/\sqrt{2}$.
Hereafter, each qubit of the quantum channel is denoted by $A$ and $B$,
respectively, and the qubit of the environment is denoted by $E$.
When the initial state of the environment is denoted by
$|0\rangle_E$, the initial total state is
\begin{equation}
\varrho_{ABE}=P^+_{AB} \otimes |0\rangle_{EE} \langle 0|,
\end{equation}
where $P^+_{AB}=|\phi^+\rangle \langle \phi^+|$.
Then, we assume that only one particle of the channel (say Alice's qubit) is
subject to the interaction with the environment.
Any type of the interaction is described by the SU(4) unitary
matrix acting on Alice's qubit and environment.
Therefore, the total state after the interaction is given by
\begin{equation}
\varrho'_{ABE}=U_{AE} (P^+_{AB} \otimes |0\rangle_{EE}\langle 0|) U_{AE}^{\dagger}.
\end{equation}
\par
The quantum channel is decohered due to the interaction,
and the reduced density matrix of the pair of $AB$ is written in
the Stinespring form \cite{Stinespring55a} as
\begin{eqnarray}
\varrho'_{AB}&=&\hbox{Tr}_E \varrho'_{ABE} 
= \sum_k (M_k \otimes 1) P^+_{AB} (M_k^\dagger \otimes 1) \cr
&\equiv& (\Lambda \otimes 1) P^+_{AB},
\end{eqnarray}
where
\begin{equation}
M_k=_E\!\langle k| U_{AE} |0\rangle_E,
\label{eq: M}
\end{equation} and $\sum_k M_k^\dagger M_k\!=\!1$.
In the same manner, the reduced density matrix of the pair of $BE$ is written
as
\begin{eqnarray}
\varrho'_{EB}&=&\hbox{Tr}_A \varrho'_{ABE}
= \sum_k (N_k \otimes 1) P^+_{AB} (N_k^\dagger \otimes 1) \cr
&\equiv& (\Gamma \otimes 1) P^+_{AB},
\end{eqnarray}
where
\begin{equation}
N_k=_A\!\langle k| U_{AE} |0\rangle_E,
\label{eq: N}
\end{equation}
and $\sum_k N_k^\dagger N_k\!=\!1$.
\par
It should be noted here that the above model has two considerable properties.
First is related to the recent work by Coffman, Kundu,
and Wootters \cite{Coffman00a}, where a relation in distributing
entanglement among three qubits was shown.
Since $\varrho'_{ABE}$ is a pure state, applying the relation directly to our
model, we obtain
\begin{equation}
C_{AB}^2+C_{EB}^2+\tau_{ABE}=C^2_{B(AE)}=1,
\label{eq: Distributed Entanglement}
\end{equation}
where $C_{AB}$ and $C_{EB}$ is the concurrence between $A$ and
$E$, and between $B$ and $E$, respectively.
$\tau_{ABE}$ is the three-qubit entanglement of the triple of $ABE$, and
$C_{B(AE)}$ is the concurrence between $B$ and the pair of $AE$.
Since $U_{AE}$ acts only on the pair of $AE$, the transformation preserves
$C_{B(AE)}$, which is equal to the initial entanglement between $A$ and $B$,
that is unity.
Therefore, in our model, $U_{AE}$ plays a role to distribute 
the entanglement initially shared by Alice and Bob to two-qubit entanglement
($AB$ and $BE$) and three-qubit entanglement.
\par
Second is a special relation established between $\varrho'_{AB}$ and
$\varrho'_{EB}$.
Choi \cite{Choi75a} and Horodecki family \cite{Horodecki99a} showed an
isomorphism between completely positive
trace-preserving (CPTP) maps and quantum states with one of subsystems being
completely mixed.
Since both $\varrho'_{AB}$ and $\varrho'_{EB}$ is obtained by CPTP map
from $P^+_{AB}$, the map $\Lambda$ and the sate $\varrho'_{AB}$ is isomorphic,
and $\Gamma$ and $\varrho'_{EB}$ is also isomorphic.
When some $\varrho'_{AB}$ is given, the map $\Lambda$ is uniquely determined
due to the isomorphism.
Although the Stinespring form of the given $\Lambda$ is not unique,
any two Stinespring forms of the same $\Lambda$ are related as
$M^{(1)}_k\!=\!\sum_\mu U_{k\mu} M^{(2)}_\mu$ with $U$ being unitary.
Further, from Eq.\ (\ref{eq: M}) and Eq.\ (\ref{eq: N}), the matrix elements
of $M$ and $N$ are related to each other through 
$[N_k]_{ij}\!=\![M_i]_{kj}$.
Therefore, two Stinespring forms of $\Gamma$ determined by $M^{(1)}_k$ and
$M^{(2)}_k$ are related as $N^{(1)}_k\!=\!U N^{(2)}_k$, where $U$ is local
unitary acting on $E$.
As a result, for some given $\varrho'_{AB}$, all the corresponding
$\varrho'_{EB}$'s are in a locally equivalent class.
\par
\begin{figure}
\epsfxsize=6.5cm
\centerline{\epsfbox{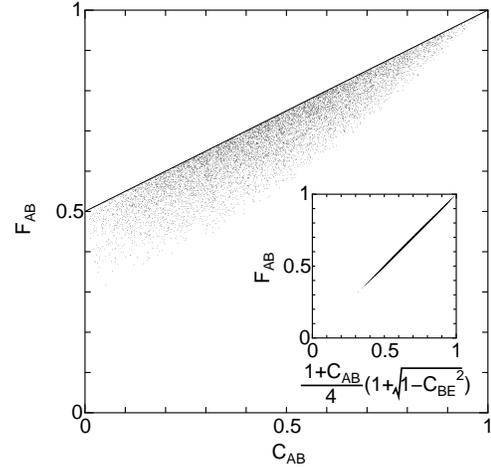}}
\caption{
Numerically obtained singlet fraction ($F_{AB}$) and
concurrence ($C_{AB}$) in our model (both are dimensionless).
A solid line shows $F_{AB}\!=\!(C_{AB}+1)/2$.
The inset shows $F_{AB}$ as
a function of $(1+C_{AB})(1+\protect\sqrt{1-C_{BE}^2})/4$.
}
\label{fig: FAB}
\end{figure}

Figure \ref{fig: FAB} shows the singlet fraction ($F_{AB}$) and the
concurrence ($C_{AB}$) of $\varrho'_{AB}$, which are obtained numerically for
10 000 random $U_{AE}$'s in the circular unitary ensemble \cite{Zyczkowski94a}.
Since $\varrho'_{AB}$ is mixed states, the points broadly distribute in the region
of $F_{AB}\!\le\!(C_{AB}+1)/2$, and we cannot see any unique relation between
$F_{AB}$ and $C_{AB}$.
However, since all $\varrho'_{EB}$'s corresponding to $\varrho'_{AB}$
are in a locally equivalent class as mentioned above, 
it is expected that some nonlocal properties of $\varrho'_{AB}$ is
transfered to $\varrho'_{EB}$, and $F_{AB}$ might be uniquely
determined by $C_{AB}$ with the help of some nonlocal property of
$\varrho'_{EB}$, as it will be shown below.
\par
In order to see this, we first examine an example for a simple case of
$U_{AE}$ as follows:
\begin{equation}
\left\{
\begin{array}{rcl}
|00\rangle_{AE} &\longrightarrow & \sqrt{1-q} |00\rangle_{AE}
                                  +\sqrt{q}|11\rangle_{AE} \cr
|10\rangle_{AE} &\longrightarrow & \sqrt{1-p} |10\rangle_{AE}
                                  +\sqrt{p}|01\rangle_{AE}. \cr
\end{array}
\right.
\end{equation}
When $q\!=\!0$, this transformation corresponds to the usual amplitude damping
for Alice's qubit.
The reduced density matrix of $\varrho'_{AB}$ are easily obtained as
\begin{eqnarray}
\varrho'_{AB}&=&
 \frac{1-q}{2}|00\rangle \langle 00|+\frac{p}{2}|01\rangle \langle 01|
+\frac{q}{2}|10\rangle \langle 10|  \cr
&+&\frac{1-p}{2}|11\rangle \langle 11|
+\Bigg[
\frac{\sqrt{(1-p)(1-q)}}{2}|00\rangle \langle 11| \cr
&&+\frac{\sqrt{pq}}{2}|01\rangle \langle 10| + \hbox{h.c.}\Bigg],
\end{eqnarray}
and the singlet fraction and concurrence of $\varrho'_{AB}$ is
\begin{eqnarray}
C_{AB}&=&| \sqrt{(1-p)(1-q)} - \sqrt{pq} |, \cr
F_{AB}&=& \frac{2-p-q+2\sqrt{(1-p)(1-q)}}{4}.
\label{eq: CAB}
\end{eqnarray}
Further, we obtain 
\begin{eqnarray}
\varrho'_{EB}&=&
 \frac{1-q}{2}|00\rangle \langle 00|+\frac{1-p}{2}|01\rangle \langle 01|
+\frac{q}{2}|10\rangle \langle 10|  \cr
&+&\frac{p}{2}|11\rangle \langle 11|
+\Bigg[
\frac{\sqrt{p(1-q)}}{2}|00\rangle \langle 11| \cr
&&+\frac{\sqrt{(1-p)q}}{2}|01\rangle \langle 10| + \hbox{h.c.}\Bigg],
\end{eqnarray}
whose concurrence is
\begin{equation}
C_{EB}=| \sqrt{(1-p)q} - \sqrt{p(1-q)} |.
\label{eq: CEB}
\end{equation}
Combining Eq.\ (\ref{eq: CAB}) and Eq.\ (\ref{eq: CEB}), we finally obtain a
rather simple relation of
\begin{equation}
F_{AB}=\frac{1+C_{AB}}{4} \Bigg(1+\sqrt{1-C_{EB}^2}\Bigg).
\label{eq: Fully Entangled Fraction}
\end{equation}
\par
Although the above relation was obtained for a special case of $U_{AE}$,
we found numerically that the relation holds for any form of $U_{AE}$.
In fact, $F_{AB}$ numerically obtained for random $U_{AE}$'s linearly depends
on $(1+C_{AB})(1+\sqrt{1-C_{EB}^2})/4$ as shown in the inset of
Fig.\ \ref{fig: FAB}.
Since all quantities in Eq.\ (\ref{eq: Fully Entangled Fraction}) are
invariant under any local unitary,
Eq.\ (\ref{eq: Fully Entangled Fraction}) holds for any maximally entangled
state as an initial state of $AB$.
It is nontrivial that the above simple relation holds, since
$U_{AE}$ has $15\!-\!2\!\times\!3\!=\!9$ independent nonlocal parameters.
Then, we arrive at the main result of this report:
{\it When an ideal quantum channel (any maximally entangled state) is
locally decohered due to any
interaction with two-level environment (another qubit), the singlet
fraction of the decohered quantum channel is determined uniquely by the
distributed entanglement as Eq.\ (\ref{eq: Fully Entangled Fraction}).}
\par
It should be noted here that, when the initial state of $AB$ is partially
entangled pure state, we could not find any unique relation.
This may be because the special relation between $\varrho'_{AB}$ and
$\varrho'_{EB}$, which is not established for
partially entangled initial states, plays a crucial role.
\par
According to Eq.\ (\ref{eq: Fidelity}) by Horodecki \cite{Horodecki99a},
when the decohered quantum channel $\varrho'_{AB}$ is used under the STS,
the optimal fidelity is given by
\begin{equation}
f=\frac{1}{2}+\frac{\sqrt{1-C_{EB}^2}}{6}+\frac{1+\sqrt{1-C_{EB}^2}}{6}C_{AB}.
\label{eq: Our Fidelity}
\end{equation}
Let's consider the physical meaning of $f$ in the STS instead of $F_{AB}$
itself.
\par
\begin{figure}
\epsfxsize=4.5cm
\centerline{\epsfbox{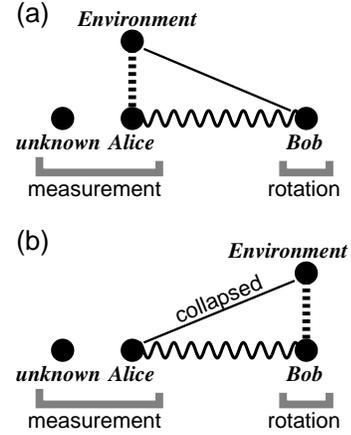}}
\caption{
When the decohered quantum channel in our model is used under the STS,
two types of configuration are present.
(a) Alice's qubit only and (b) Bob's qubit only is coupled with the
environment.
}
\label{fig: Configuration}
\end{figure}

The unknown state to be teleported is assumed to be pure state of
$\varrho_u\!=\!( 1+ \vec s \cdot \vec \sigma )/2$
for simplicity ($|\vec s|\!=\!1$).
In the STS, Alice performs a collective measurement for Alice's qubit and an
unknown state (see Fig.\ \ref{fig: Configuration} (a)).
Alice obtains the result of the measurement $\alpha$ with a probability
$P_\alpha$.
After the measurement, the state of the pair of $EB$ becomes a pure state
$|\psi_\alpha\rangle$, which depends on $\alpha$.
This represents a decomposition of the mixed state $\varrho'_{EB}$ into pure
states:
\begin{equation}
\varrho'_{EB}=\sum_{\alpha} P_\alpha |\psi_\alpha\rangle \langle \psi_\alpha |.
\end{equation}
Bob's qubit is still entangled with the environment, but the amount of the
entanglement also depends on $\alpha$.
When the concurrence of $|\psi_\alpha\rangle$ is denoted by $C_{EB}^\alpha$,
convexity of the concurrence implies
\begin{equation}
C_{EB}\le\sum_\alpha P_\alpha C_{EB}^\alpha.
\label{eq: Convexity}
\end{equation}
Then, Bob rotates the state of his qubit depending on $\alpha$, and
$|\psi_\alpha\rangle$ becomes $|\phi_\alpha\rangle$.
However, since the rotation is local in the STS, $C_{EB}^\alpha$ does not
change.
Finally, Bob obtains the state of
$\varrho_B^\alpha\!=\!\hbox{Tr}_E |\phi_\alpha\rangle\langle\phi_\alpha|$,
whose eigenvalues are
\begin{equation}
\{\frac{1}{2}+\frac{q_\alpha}{2},\frac{1}{2}-\frac{q_\alpha}{2}\}.
\label{eq: Eigenvalues}
\end{equation}
Since $|\phi_\alpha\rangle$ is a pure state, 
the eigenvalues of $\varrho_B^\alpha$ are related to $C_{EB}^\alpha$ as
\begin{equation}
q_\alpha=\sqrt{1-C_{EB}^{\alpha2}}.
\end{equation}
\par
When the quantum channel is absent and only classical communications are
allowed, it will be optimal that Alice and Bob adopt the following strategy:
Alice performs an orthogonal measurement in a spin direction (say $z$-axis)
on the unknown state and Bob prepares a state depending on the result of
Alice's measurement.
Alice measures the spin ``up'' and ``down'' with probability 
$p_\uparrow\!=\!(1+s_z)/2$ and
$p_\downarrow\!=\!(1-s_z)/2$, respectively.
In this strategy, if Bob's particle is in a pure state,
Bob can prepare the state
\begin{equation}
\varrho_B=\frac{p_\uparrow}{2}[1+\sigma_z]+\frac{p_\downarrow}{2}[1-\sigma_z]
=\frac{1}{2}[1+s_z \sigma_z].
\end{equation}
However, since Bob's particle must be entangled with the environment, Bob's
particle must be in the mixed state with eigenvalues
of Eq.\ (\ref{eq: Eigenvalues}) for each result of Alice's measurement
``up'' and ``down''.
Under this constraint, Bob can only prepare
\begin{eqnarray}
\varrho_B^\alpha=
\frac{p_\uparrow}{2}[1+q_\alpha\sigma_z]
+\frac{p_\downarrow}{2}[1-q_\alpha\sigma_z]
=\frac{1}{2}[1+q_\alpha s_z \sigma_z].
\end{eqnarray}
Therefore, using classical communications only, the attainable fidelity
averaged over $\vec s$ and $\alpha$ is
\begin{eqnarray}
f^{\rm CC}&=&\sum_\alpha P_\alpha(\frac{1}{2}+\frac{q_\alpha}{6})
=\sum_\alpha P_\alpha(\frac{1}{2}+\frac{\sqrt{1-C_{BE}^{\alpha2}}}{6}) \cr
&\le& \frac{1}{2}+\frac{1}{6}\sqrt{1-(\sum_\alpha P_\alpha C_{BE}^\alpha)^2}\cr
&\le& \frac{1}{2}+\frac{\sqrt{1-C_{BE}^2}}{6} \equiv f^{\rm CC}_{\rm max}
\end{eqnarray}
Here, we have used the concavity of $\sqrt{1-x^2}$ for the first inequality
and convexity of the concurrence Eq.\ (\ref{eq: Convexity}) for the second
inequality.
The upper bound of $f^{\rm CC}_{\rm max}$ agrees with 
Eq.\ (\ref{eq: Our Fidelity}) for $C_{AB}\!=\!0$.
\par
When an ideal quantum channel is shared by Alice and Bob and
if Bob's particle is in a pure state,
Bob can prepare $\varrho_u$ faithfully.
However, Bob's particle must be in the mixed state as discussed above.
The optimal state under the constraint is thus
\begin{equation}
\varrho_B^\alpha=\frac{1}{2}[1+ q_\alpha \vec s \cdot \vec \sigma].
\end{equation}
The fidelity averaged over $\vec s$ and $\alpha$ is again
\begin{eqnarray}
f^{\rm QC}=\sum_\alpha P_\alpha(\frac{1}{2}+\frac{q_\alpha}{2})
\le \frac{1}{2}+\frac{\sqrt{1-C_{BE}^2}}{2} \equiv f^{\rm QC}_{\rm max}.
\end{eqnarray}
As a result, the fidelity in our model Eq.\ (\ref{eq: Our Fidelity}) is
rewritten as
\begin{eqnarray}
f=f_{\rm max}^{\rm CC}+\frac{f_{\rm max}^{\rm QC}}{3}C_{AB}.
\label{eq: Meaning of Fidelity}
\end{eqnarray}
The meaning of the factor $1/3$ is that
a half of the fidelity is gained by the maximally mixed state,
and a remaining half is gained by the preparation,
but one of three degrees of $\vec s$ is already used in $f_{\rm max}^{\rm CC}$.
Therefore, $1/2\times 2/3=1/3$ of $f_{\rm max}^{\rm QC}$ is carried by the
entanglement between Alice and Bob, which linearly depends on the concurrence
$C_{AB}$.
The above expression Eq.\ (\ref{eq: Meaning of Fidelity}) can
be regarded as the natural extension of Eq.\ (\ref{eq: Pure State Fidelity})
considering the remaining entanglement between teleported state and the
environment.
In fact, Eq.\ (\ref{eq: Meaning of Fidelity}) completely agrees with
Eq.\ (\ref{eq: Pure State Fidelity}) for $C_{EB}\!=\!0$.
It is interesting to note that the fidelity is determined by the upper bound
of $f^{\rm CC}$ and $f^{\rm QC}$.
In this sense, the STS seems to be optimal under the constraint of the
remaining entanglement, at least in our model.
\par
In the above, we consider the case that Alice's qubit is coupled with
the environment.
When Bob's qubit is coupled with the environment as shown in 
Fig.\ \ref{fig: Configuration} (b), $C_{EB}$ in 
Eq.\ (\ref{eq: Fully Entangled Fraction}) must be read as $C_{EA}$.
In this case also, Bob's qubit is entangled with the environment after
the procedure of STS.
However, the state of the pair of $EB$ has no special relation
to the state of the quantum channel.
The state $\varrho'_{EA}$, which has a special relation,
is collapsed by Alice's measurement.
Therefore, it is difficult to discuss the physical meaning when Bob's qubit
is coupled with the environment.
Further, in this configuration, only an inequality
\begin{eqnarray}
f&=&\frac{3+\sqrt{1-C_{AE}^2}}{6}+\frac{1+\sqrt{1-C_{AE}^2}}{6}C_{AB} \cr
&\le&\frac{3+\sqrt{1-C_{EB}^2}}{6}+\frac{1+\sqrt{1-C_{EB}^2}}{6}C_{AB}
\end{eqnarray}
is satisfied, since we numerically confirmed that $C_{AE}\!\ge\!C_{BE}$ 
in our model, though results are not shown explicitly here.
In this sense, the STS for the configuration of Fig.\ \ref{fig: Configuration}
(b) might be less optimal than that of Fig.\ \ref{fig: Configuration} (a)
under the constraint of the remaining entanglement.
\par
It is important to note that the teleportation fidelity itself is the
same in both configurations of Fig.\ \ref{fig: Configuration} (a) and (b),
since $F_{AB}$ is the same independent of the configuration.
In this sense, the expression of $F_{AB}$ should be symmetric under the
exchange of $A$ and $B$.
For this purpose, Eq.\ (\ref{eq: Fully Entangled Fraction}) can be rewritten
by using Eq.\ (\ref{eq: Distributed Entanglement}) as 
\begin{equation}
F_{AB}=\frac{1+C_{AB}}{4} \Bigg(1+\sqrt{C_{AB}^2+\tau_{ABE}}\Bigg),
\label{eq: Symmetric Fully Entangled Fraction}
\end{equation}
which is symmetric since $\tau_{ABE}$ is symmetric \cite{Coffman00a}.
However, since three-qubit entanglement $\tau_{ABE}$ is also collapsed by the
Alice's measurement, it will be hard to discuss the physical meaning
in the STS.
\par
In this report, we exclusively consider the case that the environment is a
two-level system.
In order to discuss general local decoherence of the quantum channel, at
least four-level environment must be considered.
Even in this case, the special relation between states with respect
to the nonlocal properties,
which plays a crucial role in this report, is also established.
However, the measure of the entanglement for the 2$\times$4 system is
necessary for this purpose, and it is important to clarify the nature of
entanglement in such larger dimensional systems.

\end{document}